\documentclass[11pt]{article}
\usepackage{graphicx,amsmath,amsthm,mathrsfs,amssymb}
\usepackage[usenames]{color}
\usepackage{ulem}

\newcommand{\ee}{\end{equation}}
\newcommand{\word}[1]{\,\,\mbox{#1}\,\,}
\newcommand{\reff}[1]{(\ref{#1})}
\newcommand{\beq}{\begin{equation}}
\newcommand{\eeq}[1]{\label{#1}\end{equation}}
\newcommand{\beqa}{\begin{eqnarray}}
\newcommand{\eea}{\end{eqnarray}}
\newcommand{\eeqa}[1]{\label{#1}\end{eqnarray}}
\newcommand{\beg}{\begin{equation*}}
\newcommand{\eeg}{\end{equation*}}

\newcommand{\eq}{\!=\!}
\newcommand{\p}{\!+\!}
\newcommand{\m}{\!-\!}

\newcommand{\less}{\!<\!}
\newcommand{\great}{\!>\!}

\newcommand{\abs}[1]{\lvert#1\rvert}

\newcommand{\bsplit}{\begin{split}}
\newcommand{\esplit}{\end{split}}

\usepackage[usenames]{color}
\usepackage{ulem}
\begin{document}
\begin{titlepage}
\title{Emergence of a thin shell structure \\during collapse in isotropic coordinates}
\author{\thanks{hbeauchesne10@ubishops.ca} Hugues Beauchesne and \thanks{aedery@ubishops.ca} Ariel Edery \\\\{\small\it Physics Department, Bishop's University}\\
{\small\it 2600 College Street, Sherbrooke, Qu\'{e}bec, Canada
J1M~0C8}}
\date{}
\maketitle
\begin{abstract}

Numerical studies of gravitational collapse in isotropic coordinates have recently shown an interesting connection between the gravitational Lagrangian and black hole thermodynamics. A study of the actual spacetime was not the main focus of this work and in particular, the rich and interesting structure of the interior has not been invetigated in much detail and remains largely unknown. We elucidate its features by performing a numerical study of the spacetime in isotropic coordinates during gravitational collapse of a massless scalar field. The most salient feature to emerge is the formation of a thin shell of matter just inside the apparent horizon. The energy density and Ricci scalar peak at the shell and there is a jump discontinuity in the extrinsic curvature across the apparent horizon, the hallmark that a thin shell is present in its vicinity. At late stages of the collapse, the spacetime consists of two vacuum regions separated by the thin shell. The interior is described by an interesting collapsing isotropic universe. It tends towards a vacuum (never reaches a perfect vacuum) and there is a slight inhomogeneity in the interior that plays a crucial role in the collapse process as the areal radius tends to zero. The spacetime evolves towards a curvature (physical) singularity in the interior, both a Weyl and Ricci singularity. In the exterior, our numerical results match closely the analytical form of the Schwarzschild metric in isotropic coordinates, providing a strong test of our numerical code.  
 
\end{abstract}
\thispagestyle{empty}
\end{titlepage}
\setcounter{page}{2}
\section{Introduction}

Isotropic coordinates have recently been used to study numerically black hole thermodynamics during gravitational collapse \cite{Khlebnikov, Benjamin-Edery2} and to study numerically the evolution of metric perturbations during reheating \cite{Finelli}. These coordinates are convenient for numerical studies of gravitational collapse because the metric functions are not singular anywhere \cite{Khlebnikov, Finelli}. In particular, the functions are continuous and finite across the horizon and the entire spacetime is covered with one coordinate patch. The metric is in a $3+1$ decomposition form allowing for a Hamiltonian or Lagrangian formulation \cite{Poisson}. Moreover, the coordinate $t$ represents the time measured by an asymptotic observer at rest and this feature allows one to relate the gravitational Lagrangian to a thermodynamic potential making isotropic coordinates ideal for black hole thermodynamic studies \cite{Khlebnikov}. In particular, it was shown numerically that in isotropic coordinates the negative of the gravitational Lagrangian approaches the free energy of a black hole at late stages of collapse \cite{Khlebnikov, Benjamin-Edery2}.

In a $3+1$ decomposition \cite{Poisson} a metric which is spherically symmetric takes on the following general form:
\beq
ds^2\eq -(N^2 - A^2\beta^2)\,dt^2 +2\,A^2\beta \,dr dt + A^2\,dr^2 +B^2\,r^2d\Omega^2 \,.
\eeq{Spherical}
There are altogether four metric functions of $r$ and $t$: the lapse $N(r,t)$, the shift $\beta(r,t)$ and two functions $A(r,t)$ and $B(r,t)$ for the spatial three-metric. We can use gauge freedom, the twofold ambiguity associated with choosing two different coordinates, to reduce the number to two metric functions. The isotropic coordinate system is based on the choice $\beta=0$ and ``isotropic gauge" $A^2=B^2 =\psi^4$ for the spatial metric so that \reff{Spherical} reduces to  
\beq
ds^2= -N^2(r,t)\,dt^2 +\psi^4(r,t) (dr^2 + r^2 d\Omega^2)\,. 
\eeq{isometric}
$\psi$ is called the conformal factor. The spherically symmetric vacuum solution in isotropic coordinates is given analytically by \cite{Wald}
\beq
ds^2=-\dfrac{(1-GM/2r)^2}{(1+ GM/2r)^2}\,dt^2 +\Big(1+\dfrac{GM}{2r}\Big)^4(dr^2+r^2\,d\Omega^2)\,.
\eeq{isotropic}
The event horizon is located at $r\eq GM/2$ and the region $r\great GM/2$ covers the exterior static region of the Schwarzschild black hole. However, the region $r\less GM/2$ does not cover the interior of the Schwarzschild black hole; it covers the static exterior region a second time so that metric \reff{isotropic} is a double covering of the Schwarzschild exterior. In standard coordinates, the Schwarzschild metric has the form $ds^2\eq  -A(R)\,dt^2 + B(R)\,dR^2 + R^2 \,d\Omega^2$ where $A(R)\eq  1-2GM/R$ and $B(R)\eq A(R)^{-1}$. The coordinate transformation $R\eq r\,(1+ \tfrac{GM}{2r})^2$ yields the metric in the isotropic form \reff{isotropic}. We see that $r\eq GM/2$ corresponds to the event horizon $R\eq 2GM$. The regions $r\great GM/2$ and $r\less GM/2$ both correspond to $R\great 2GM$, the exterior Schwarzschild region. The interior region $R\less 2GM$ of the Schwarzschild black hole is not covered by metric \reff{isotropic}. This is clear when we consider the Killing vectors. In the interior region of the Schwarzschild black hole, all Killing vectors are spacelike \cite{Edery_Constantineau} and this implies that the interior is nonstationary. In contrast, the metric \reff{isotropic} has a timelike hypersurface-orthogonal Killing vector in the region $r\less GM/2$ and is therefore static in that region. To explore the interior in isotropic coordinates, the metric must be of the time-dependent form \reff{isometric}. During collapse, the metric \reff{isometric} is not singular at the horizon: the functions $N$ and $\psi$ are continuous across the apparent horizon and finite everywhere. Moreover, the coordinate $t$ remains timelike in the interior. Note that the coordinate $t$ in isotropic coordinates represents the time measured by an asymptotic observer at rest. Such a time coordinate has not been used in most previous numerical studies of the interior during spherically symmetric collapse. This includes scalar field collapse in both Painlev\'{e}-Gullstrand (PG) coordinates \cite{Kunstatter} and in coordinates used for studying the evolution of trapped-surfaces \cite{Istvan}, Einstein-Yang Mills collapse using maximal slicing and radial spatial coordinates \cite{Marsa} and charged scalar field collapse in ``Eddington-Finkelstein" \cite{Brady} or double-null coordinates \cite{Piran}.  

Numerical studies of gravitational collapse in isotropic coordinates have largely focused on the thermodynamics \cite{Khlebnikov, Benjamin-Edery2} and to date, a separate analysis of the interior spacetime is lacking. The interior in isotropic coordinates has a rich structure and we elucidate it by studying numerically the entire spacetime during the gravitational collapse of a massless scalar field to a black hole. The most salient feature is the emergence of a thin shell of matter at a radius just below $r_h$, the radial location of the apparent horizon (where the lapse function $N(r,t)$ is zero at late times). The energy density peaks just inside the apparent horizon (and hence inside the event horizon) and the extrinsic curvature undergoes a jump discontinuity across the horizon, the hallmark that a thin shell has formed in its vicinity \cite{Poisson}. By tracking curvature scalars, the spacetime can be seen to be evolving towards a curvature singularity in the interior (both a Weyl and Ricci singularity). The region outside the horizon $r>r_h$ evolves towards the metric \reff{isotropic}, the static vacuum Schwarzschild exterior in isotropic coordinates. The interior tends towards a vacuum and corresponds to a collapsing isotropic universe where the trace of the extrinsic curvature is spatially constant and increases rapidly with time. The interior is not exactly a Friedmann-Lema\^{i}tre-Robertson-Walker (FLRW) universe because it is slightly inhomogeneous. The small inhomogeneity plays a significant role in the collapse process as $\psi\to 0$. The areal radius is given by $R=\psi^2 r$ where $r$ is the radial coordinate in isotropic coordinates. In the interior, $\psi$ ranges from $0$ to a value of $2$ at $r=GM/2$ so that the areal radius ranges from $0$ to $2M$ in the interior; the entire spacetime, both interior and exterior, is covered in isotropic coordinates. In the work of \cite{Khlebnikov,Benjamin-Edery2}, both the exterior and interior region make a contribution to the free energy and the entire spacetime is needed to obtain numerically the thermodynamics for the Schwarzschild black hole.

Radial null geodesics in isotropic coordinates have $dr/dt$ equal to zero at the $N\eq0$ two-surface (where $r$ and $t$ are the radial and time coordinate in isotropic coordinates). Matter accumulates in a thin shell near the $N\eq 0$ surface, which at late times is situated at a finite (non-zero) radius $r_h$, the radius of the apparent horizon. The thin shell forms just inside the apparent horizon and is seen by a static observer in the interior at $r\eq 0$ (note that a static observer at $r\eq 0$ remains static at all times). The proper radial distance between $r=0$ and $r=r_h$ tends to zero with time so that the thin shell is ``falling" towards $r\eq 0$ even though it remains at a radial location near $r_h$. The proper radial distance is given by $\int_0^{r_h}\psi^2(r,t)\,dr$ which is approximately equal to $\psi^2(t) \,r_h$ (since $\psi$ is approximately homogeneous in the interior). This tends towards zero as $\psi$ approaches $0$. 

Analytical studies have shown that when matter is thrown into a Schwarzschild black hole, a {\it static asymptotic observer} sees the matter form into thin ``pancakes" at the horizon \cite{Lindesay}. The formation of thin layers of matter at the horizon is not seen by freely falling observers (called by the acronym Frefos in \cite{Lindesay}). In isotropic coordinates, the time $t$ is the time measured by a static asymptotic observer. The formation of a thin shell at a radius near the horizon is the dynamical realization of the thin ``pancake" scenario discussed in \cite{Lindesay}.  As far as we know, our work is the first time that a numerical collapse study has explicitly shown and discussed this. Moreover, our numerical and hence dynamical results yield new insights on this phenomena. An analytical test-particle approximation would conclude that the ``thin pancake" remains just {\it outside} the horizon. Our numerical simulation shows that the matter accumulates in a thin shell just {\it inside} the horizon, not outside. To observe this analytically would require one to include the back-reaction. The numerical simulations of spherically symmetric collapse previously mentioned \cite{Kunstatter}-\cite{Piran}, do not observe the thin shell at the horizon discussed in \cite{Lindesay}. This is not surprising since these studies do not use a coordinate corresponding to the time measured by a static observer at spatial infinity.  

The viewpoint of the static asymptotic observer is important for understanding black hole thermodynamics. The temperature $T=1/(8\pi M)$ of a Schwarzschild black hole is the temperature as seen by an observer at rest at spatial infinity \cite{Lindesay} and its entropy represents a measure of an external observer's ignorance of the internal configurations hidden behind the event horizon \cite{Bekenstein, Edery_Constantineau}. The formation of a thin shell near the horizon in isotropic coordinates plays a significant role in black hole thermodynamics and the holographic principle. In thermodynamic studies \cite{Khlebnikov, Benjamin-Edery2}, it was observed that the black hole entropy is concentrated in a thin shell just inside the horizon. Isotropic coordinates offers a vantage point from where one can observe the entropy accumulate directly on the surface of the horizon and this represents a direct and dynamical observation of the holographic principle \cite{tHooft,Susskind}.

Our paper is organized as follows. In section 2 we state the equations of motion, initial states and boundary conditions that are used in our numerical simulation (these are derived in detail in \cite{Benjamin-Edery2}). In section 3 we discuss the formation of the black hole. In particular, we state the defining property of the $N\eq 0$ spacelike two-surface and discuss how matter enters into the interior as the $N\eq 0$ surface expands (and is identified as the apparent horizon at late times). In section 4 we analyze the entire spacetime: the interior collapsing universe, the thin shell region and the exterior vacuum region, which acts as a benchmark for our numerical simulation. 
              
\section{4D gravitational collapse of a scalar field \\in isotropic coordinates}

The spacetime during the collapse process is described by the spherically-symmetric time-dependent metric \reff{isometric}. The function $\psi(r,t)$, the conformal factor, is a dynamical variable but the lapse function $N(r,t)$ is not \footnote{The lapse changes with time. By ``not dynamical" we mean that its momentum conjugate is not defined because the gravitational action does not contain time-derivatives of $N$ \cite{Poisson}.}. Though it is not dynamical, each time step yields a new spacelike hypersurface $\Sigma_t$ so that one must reevaluate the value of $N(r,t)$ at the new surface. This is governed by an ordinary differential equation instead of an evolution equation. The Einstein field equation governing the evolution of $\psi$ contains two time-derivatives. For numerical purposes, this is broken into two single time-derivative equations. The Einstein field equations also yield two constraint equations: one for the energy and the other for momentum. These are not evolution equations but are useful for monitoring the accuracy of the simulation. The collapsing matter is a Klein-Gordon massless scalar field $\chi(r,t)$ which is minimally coupled to the gravitational field. Again, the evolution of $\chi$ is governed by two single time-derivative equations. Once the equations of motion are supplemented with initial states and boundary conditions, the evolution of the metric and matter fields is unique. Depending on the initial state, the system either collapses to a black hole or disperses due to the internal pressure of the scalar field. We are interested only in the case of black hole formation.         

\subsection{4D equations of motion}

The Einstein field equations yield an energy constraint, a momentum constraint, an evolution equation for the conformal factor $\psi$, an evolution equation for its ``momentum conjugate", the trace of the extrinsic curvature $K$ and finally an ODE for the lapse function $N$. In 4D, the equations are (see \cite{Benjamin-Edery2} for details and derivation):
\begin{align}
&\word{energy constraint:}\quad	- \frac{4}{\psi^5}\,\nabla^2\psi = \kappa^2\mathcal{E} - \frac{K^2}{3}\,. \label{E_Constraint}\\
&\word{momentum constraint:}\quad\frac{K'}{3} = \frac{\kappa^2}{2}\,\frac{\dot\chi}{N}\,\chi'
\label{M_Constraint}\\
&\word{evolution equation for} \psi:\quad	\frac{\dot{\psi}}{N} = -\frac{K\psi}{6}
\label{Psi_Evo}\\
&\word{evolution equation for K:} \quad \frac{\dot{K}}{N} = \frac{K^2}{2} (1+w) + \frac{3}{2}\kappa^2\mathcal{E} (1-w) - 6\frac{\psi'}{\psi^5} \Big( \frac{\psi'}{\psi} + \frac{1}{r} \Big)	\label{K_Evo}\nonumber\\
&\qquad\qquad\qquad\qquad\qquad\qquad\qquad\qquad - 3\frac{N'}{N\psi^4} \Big( \frac{2\psi'}{\psi} + \frac{1}{r} \Big) - 6w\frac{\nabla^2\psi}{\psi^5}\\
&\word{ODE for N:} \quad \frac{2r}{\psi^2}\partial_r \Big( \frac{\psi'}{r\psi^3} \Big) + \frac{r}{N}\partial_r \Big( \frac{N'}{r\psi^4} \Big) = -\kappa^2\frac{\chi'^2}{\psi^4}	\label{N_ODE}
\end{align}
where $\kappa^2\equiv 8\,\pi\,G$ and the arbitrary parameter $w$ in \reff{K_Evo} arises from a term added for numerical stability purposes (we set it to unity). The energy density of the matter field is defined as
\begin{equation}
			\mathcal{E}\equiv \frac{1}{2} \Big( \frac{\chi'^2}{\psi^4} + \frac{\dot{\chi}^2}{N^2} \Big).
		\label{Energy_Density}
		\end{equation}
There are also the evolution equations for the matter field $\chi$ and its ``momentum conjugate" $p$. These are given by \cite{Benjamin-Edery2} 
\beq
p \equiv \psi^6\frac{\dot{\chi}}{N}
\eeq{Chi_Evo}
and
	\beq
			\frac{\dot{p}}{N} = \psi^2 \Big( \chi'' + \frac{N'}{N}\chi' + \frac{2\psi'}{\psi}\chi' + \frac{2}{r}\chi' \Big)\,.
	\eeq{P_Evo}

\subsection{Initial states and boundary conditions}

We choose the same initial states and boundary conditions as in \cite{Benjamin-Edery2}. We will therefore be brief. We choose a static initial state, where $K$ and $p$ are zero. From \reff{Psi_Evo} and \reff{Chi_Evo}, this implies that $\dot{\psi} = \dot{\chi} = 0$ at time $t=0$. For this initial state, the momentum constraint \reff{M_Constraint} is automatically satisfied. The initial field configuration for the scalar field $\chi$ is chosen to be \cite{Benjamin-Edery2}
	\beq
		\chi(t=0,r) = \frac{8\lambda^2r^4}{(\lambda^2 + r^2)^4},
	\eeq{4DScalarIEChi}
where $\lambda$ is a scale parameter. Note that the though the scalar field configuration has one peak, the energy density profile has two peaks. For $\lambda$ sufficiently small, the self-gravitational attraction is large enough to initiate a gravitational collapse to a black hole. 
	
The initial state for the conformal factor $\psi$ is obtained by solving the energy constraint \reff{E_Constraint} with the static initial conditions ($K\eq p\eq 0$) i.e. 
	\beq
		\frac{1}{r^2} \partial_r ( r^2\psi') =  -\frac{\kappa^2}{8}\chi'^2\psi.
	\eeq{E2_Constraint}
	
Asymptotically, the spacetime is flat: $\psi \to 1$ and $N\to 1$. We therefore express $\psi$ as a power expansion in the parameter $\zeta \eq \kappa^2$ i.e. 
\beq
\psi = 1 + \zeta\psi_1 + \zeta^2\psi_2 + \ldots\,.  
\eeq{ExpPsi}
and matching the left hand side with the right hand side of \reff{E2_Constraint} order by order in $\zeta$ yields the initial state for $\psi$ \cite{Benjamin-Edery2}:
	\beqa
		\psi & \eq & 1 \p \dfrac{\zeta}{\lambda^3}\frac{1}{40320\,r\,(r^2 \p \lambda^2)^8} \Big[ 1575\,r^{15}\lambda \p 12075\,r^{13}\lambda^3 \p .. \p 1575\,(r^2 \p \lambda^2)^8 \tan^{-1}(r/\lambda) \Big] \nonumber\\  
		 && + \dfrac{\zeta^2}{\lambda^8}\frac{1}{10538886758400} \Big[ 2010133125\pi^2 - \frac{2}{r(r^2+\lambda^2)^{16}} \Big( -r\lambda^2 \big( 2967339375\,r^{30} \nonumber\\
		& &  + 47828405625\,r^{28}\lambda^2 + \ldots + 60640845881\,r^2\lambda^{28} + 4113561991\,\lambda^{30} \big) \nonumber\\
		& &  + 1276275\,\lambda ( r^2 \p \lambda^2)^8 \big( 825\,r^{16} \p 4500\,r^{14}\,\lambda^2 \p .. \p 15436\,r^2\lambda^{14} \p 2717\lambda^{16} \big) \tan^{-1}(r/\lambda) \nonumber\\
		& &  + 4020266250\,r\,(r^2 + \lambda^2)^{16} [\tan^{-1}(r/\lambda)]^2 \Big) \Big] \nonumber\\
		& &  + \ldots \nonumber
	\eeqa{Expansion_Psi}

Our expansion is carried to six orders but we only show here the first two order terms. Once the initial state of $\psi$ is obtained, the initial state for $N$ is obtained numerically by solving its associated ODE \reff{N_ODE}. We iterate backwards to obtain $N$ starting at the outer computational boundary $r\eq R$, where the spacetime is flat and $N\eq1$. 

The above equations and initial states need to be supplemented with appropriate boundary conditions at the origin $r\eq0$ and the outer computational boundary $r \eq R$. To ensure regularity of the solution at $r\eq 0$, we impose the following boundary conditions: $\chi'(0,t)\eq0$ and $K'(0,t)\eq0$. To ensure asymptotic flatness at the computational outer boundary $r\eq R$ we impose $N(R,t)\eq 1$, $K'(R,t)\eq 0$ and $p'(R,t)\eq 0$. Together with the initial matter and metric states, these lead to a unique evolution.

\section{Black hole formation}

We work in geometrized units where $G\eq c \eq1$ (energy, mass, time and distance are measured in length). 

\begin{figure}[tbp]
		\begin{center}
			\includegraphics[scale=.50, draft=false, trim=2cm 1.5cm 2.5cm 2cm, clip=true]{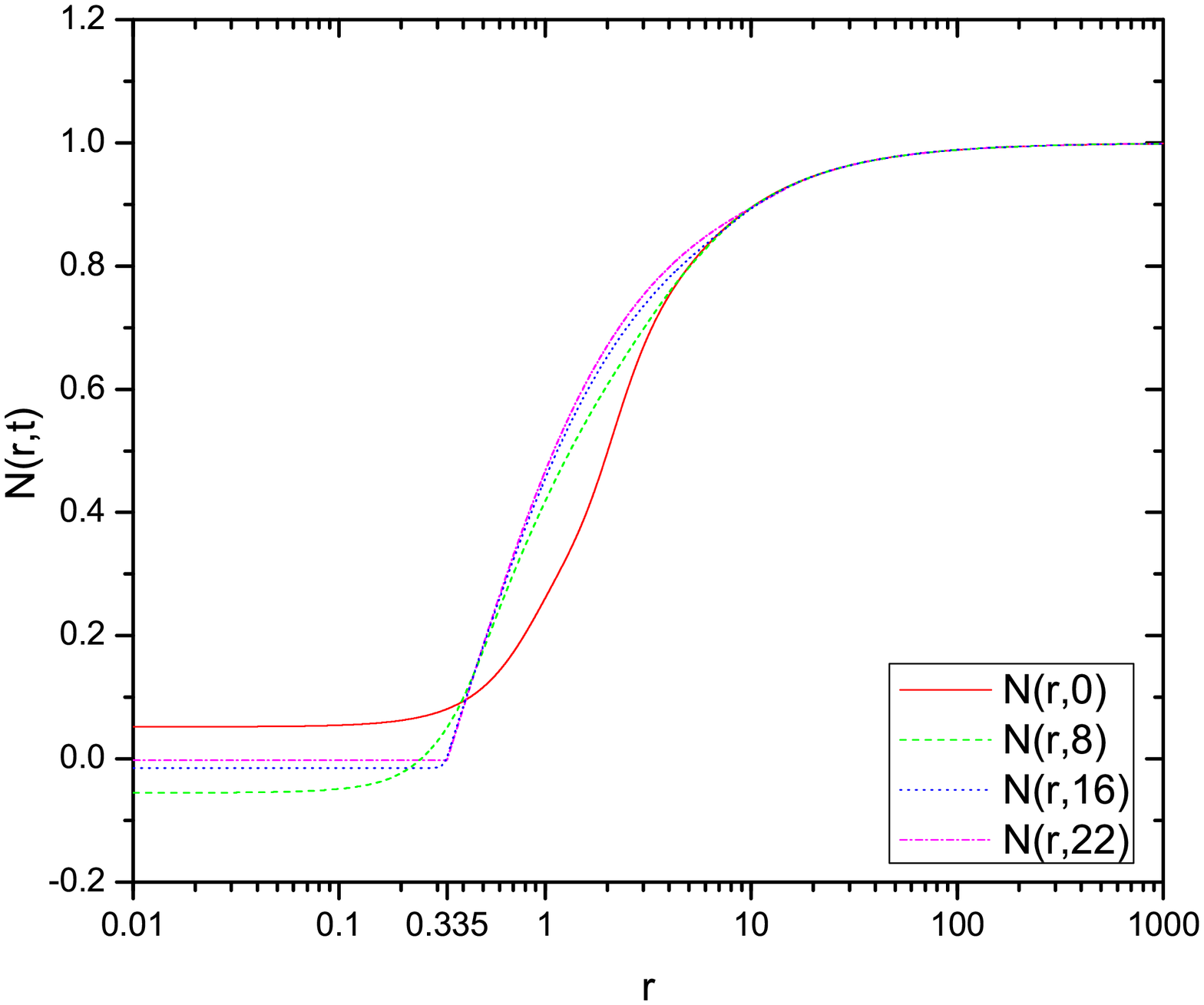}
		\end{center}
		\caption{Lapse function $N$ as a function of $r$ at different times. The radius $r_0$ where $N\eq 0$ increases with time and reaches $r_h=0.335$ at late times. The two-surface at $r_h$ is an apparent horizon. Note that $N$ is negative in the region $r<r_0$ and approaches zero from below.}
		\label{N}
\end{figure}

\begin{figure}[tbp]
		\begin{center}
			\includegraphics[scale=.50, draft=false, trim=2cm 1.5cm 2.5cm 2cm, clip=true]{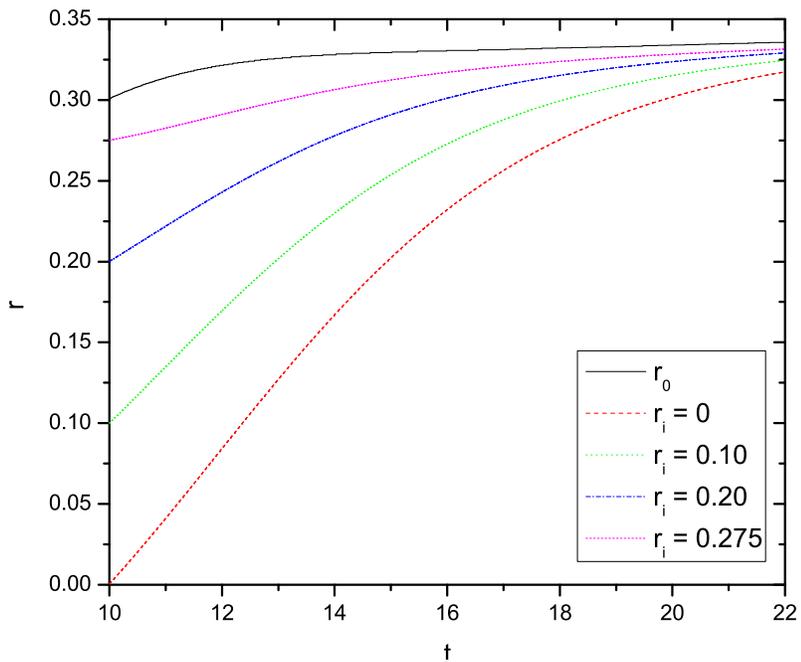}
		\end{center}
		\caption{The radius of outgoing null geodesics as a function of time. The solid line represents $r_0$. $r_i$ stands for the position of the photon particle at $t=10$. The particles start inside the $N\eq 0$ surface where $r_0=0.30$ at $t=10$. The null curves never cross the $r_0$ curve during the evolution; they remain inside the $N\eq0$ surface as both evolve.}
		\label{photons}
\end{figure}

As in \cite{Finelli}, a black hole will be considered to have formed when the lapse function $N(r,t)$ crosses zero. Let $r_0$ be the radius where $N=0$. If this radius remains fixed with time, light could not cross the $N\eq0$ two-surface in either direction since it takes an infinite amount of time $t$ for light to travel from any point to this fixed surface (as viewed of course by an asymptotic observer whose clock measures the time $t$). However, in our collapsing system, the radius $r_0$ changes as a function of time and increases until it reaches a radius of $r_h\eq 0.335$ at late times (this is the radius in figure~\ref{N}) where the lapse function crosses zero at $t\eq 22$). Note that $N$ remains negative inside once it crosses zero. The defining property of the $N\eq 0$ two-surface during the simulation is that outgoing radial null geodesics inside the surface never cross it as it evolves. The radius as a function of time for outgoing null geodesics that start inside at different radii is plotted in figure~\ref{photons}. The radius $r_0$ as a function of time is plotted as well. The null curves never intersect the $r_0$ curve; they never cross the $N\eq 0$ surface as it evolves. The null curves become flat at late times and converge towards $r_0$ reaching a radius just below $r_h=0.335$. Outgoing (and of course ingoing) null geodesics inside are trapped within this radius.  

A simple analytical argument also shows that outgoing null geodesics on the inside do not cross the $N\eq 0$ surface. From the metric, radial null geodesics obey the relation $(dr/dt)^2 = N^2/\psi^4$ and this implies that $dr/dt=0$ at the $N\eq 0$ surface (for both ingoing and outgoing null geodesics). The condition for radial null geodesics to cross the surface from inside to outside is that $dr/dt-dr_0(t)/dt$ be positive when evaluated at $r\eq r_0$. This condition is clearly not satisfied since $dr/dt=0$ at $r\eq r_0$ and $dr_0(t)/dt >0$. However, an ingoing radial null geodesic can cross the surface from outside to inside. This requires that $dr/dt-dr_0/dt$ be negative when evaluated at $r\eq r_0$ and this is automatically satisfied since again $dr/dt =0$ at $r\eq r_0$ and $dr_0/dt>0$. In particular, matter outside the surface can penetrate inside allowing the black hole mass to increase during the evolution. At late times, the $N=0$ surface is basically stationary at $r\eq r_h$ ($dr_0/dt \to 0$) and the flow of matter from outside to inside comes to a halt with the black hole mass reaching its maximum value of $M_{BH}$. 

The expansion scalar $\Theta$ for outgoing null geodesics orthogonal to the spacelike two-surface $N(r,t)=0$ (the two-sphere $ds^2=\psi^4\,r^2 d\Omega^2$) can be readily calculated \cite{Shapiro} and is equal to 

\beq
\Theta = h^{\mu\nu}\,\nabla_{\mu} k_{\nu} = 2\,\sqrt{2}\Big(\dfrac{\dot{\psi}}{\abs{N}\psi} + \dfrac{\psi'}{\psi^3}\Big) + \dfrac{\sqrt{2}}{\psi^2\,r}  
\eeq{theta}
where $h_{\mu\nu}=g_{\mu\nu} + k_{\mu}\ell_{\nu} +\ell_{\mu}k_{\nu}$ is the two-dimensional metric and $k^{\mu}= \dfrac{1}{\sqrt{2}}(1/\abs{N}, 1/\psi^2,0,0)$ and $\ell^{\mu}= \dfrac{1}{\sqrt{2}}(1/\abs{N}, -1/\psi^2,0,0)$ are future-directed null vectors orthogonal to the two-surface with normalization $k^{\mu}\ell_{\mu} =-1$. $\Theta$ is tracked numerically, and at late times, it is equal to zero at the $N\eq 0$ surface. In particular, the spacelike two-surface $r\eq r_h=0.335$ is an apparent horizon. The event horizon is a null hypersurface which is defined globally and requires knowledge of the entire future history of the spacetime. In practice, however, it can be located after a finite evolution time. During the simulation, we plotted the norm of the vector orthogonal to hypersurfaces of constant $N$. We identified a null surface at a radius $r_{null}$ outside the $N\eq 0$ surface ($r_{null}>r_0$). With time, $r_{null}$ and $r_0$ approach each other so that the event horizon is located at a radius just outside the apparent horizon ($r_h=0.335$). This is in agreement with the fact that outgoing (and of course ingoing) null geodesics inside the event horizon are trapped and cannot escape out to infinity (see fig. \reff{photons}).

\begin{figure}[tbp]
		\begin{center}
			\includegraphics[scale=.50, draft=false, trim=2cm 1.5cm 2.5cm 2cm, clip=true]{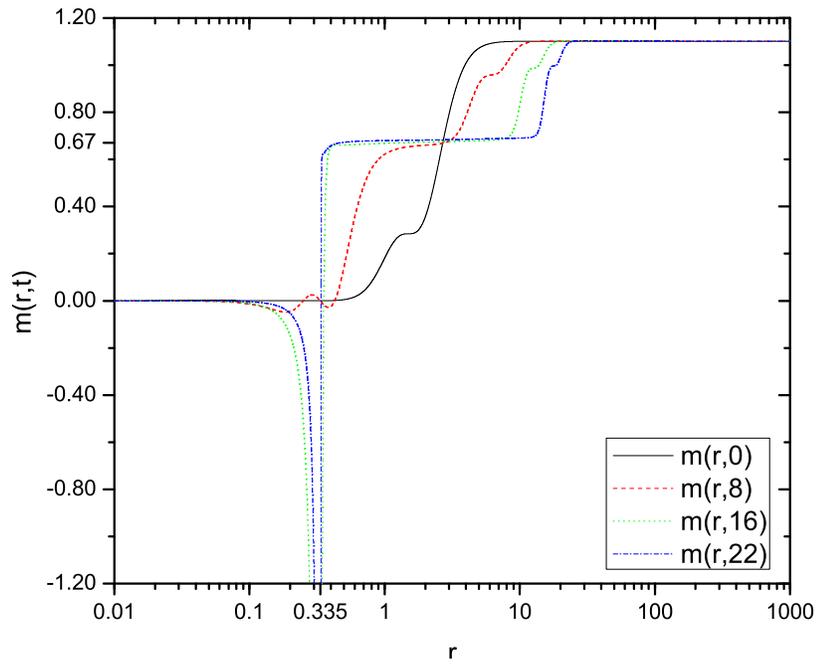}
		\end{center}
		\caption{ The mass accumulation function $m(r,t)$ as a function of $r$ at different times. The first plateau at $m \approx 0.67$ corresponds to the black hole mass $M_{BH}$ and the second plateau corresponds to the ADM mass.}
		\label{m}
\end{figure}

We know from the exterior analytical solution in isotropic coordinates that $r_h=M_{BH}/2$ where $M_{BH}$ is the black hole mass. In our dynamical collapse scenario, the black hole mass $M_{BH}$ is less than the total (conserved) ADM mass because part of the ADM mass stems from an outgoing matter wave. The function $m(r,t)\eq-r^2\,\psi'$ evaluated at infinity corresponds to the ADM mass \cite{Khlebnikov,Benjamin-Edery2}. To illustrate the difference between the black hole mass and the ADM mass, we plot $m(r,t)$ as a function of $r$ at different times in figure~\ref{m}.  Note that at large $r$, one can distinguish between two plateaus at late times. The value at the first plateau corresponds to the black hole mass $M_{BH}$  and the value at the second plateau is the ADM mass. The difference between the two values is due to an outgoing matter wave which propagates to infinity with time. The identification of the first plateau with $M_{BH}$ is confirmed by the fact that $M_{BH}/2$ at late times is equal to $r_h=0.335$ to within $2\%$. In figure~\ref{m}, the function $m(r,t)$ at late times dips to a negative value inside the horizon and then rises to reach a final positive value outside the horizon. The origin of these two contributions can be determined by expressing the function $m(r,t)$ in integral form via the energy constraint \reff{E_Constraint}:
\beq
m(r,t)= 4\,\pi\int_0^r \big(\mathcal{E}- \tfrac{K^2}{3\kappa^2}\big) \psi^5\,r'^2\,dr'\,.
\eeq{mrt}
The negative dip stems from the $-K^2$ term, a gravitational term, and the rise to a positive value stems from the positive energy density $\mathcal{E}$ of the matter. Their sum (the integral) yields the positive mass $M_{BH}$ of the black hole. This is a clear illustration of the fact that the mass of a black hole stems not only from matter but also from gravitation itself \cite{Poisson}.    

\section{Features of the spacetime: numerical results}

\subsection{Exterior vacuum region: a benchmark}
Past numerical studies \cite{Khlebnikov, Benjamin-Edery2} have shown that the collapse of matter in isotropic coordinates leads to metric and matter functions whose gradients change sharply in the region near the apparent horizon. To place more points in the interior region we expressed the radius $r$ as a function of a parameter $x$: $r=\frac{2x}{1-x}$. $x$ runs from $0$ to $1$ and $r$ runs from $0$ to ``infinity" (i.e. the outer computational boundary). At highest resolution, we used $2 \times 10^4$ points for a step size of $\Delta x=5 \times 10^{-5}$ (and $\Delta t=1 \times 10^{-4}$). We used the initial state of \reff{4DScalarIEChi} with $\lambda=1.5$. To evolve the fields in time, we used a fourth-order Adams-Bashforth-Moulton (ABM) explicit scheme. During the evolution we monitored the ADM mass i.e. $M= 4\,\pi\int_0^R \big(\mathcal{E} - \tfrac{K^2}{3\kappa^2}\big) \psi^5\,r^2\,dr$ where $R$ is the outer computational boundary, $\mathcal{E}$ is the energy density given by \reff{Energy_Density} and $K$ is the trace of the extrinsic curvature ($\kappa^2=8\pi$ since we work in geometrized units where $G\eq c \eq 1$). 

\begin{figure}[tbp]
		\begin{center}
			\includegraphics[scale=.50, draft=false, trim=2cm 1.5cm 2.5cm 2cm, clip=true]{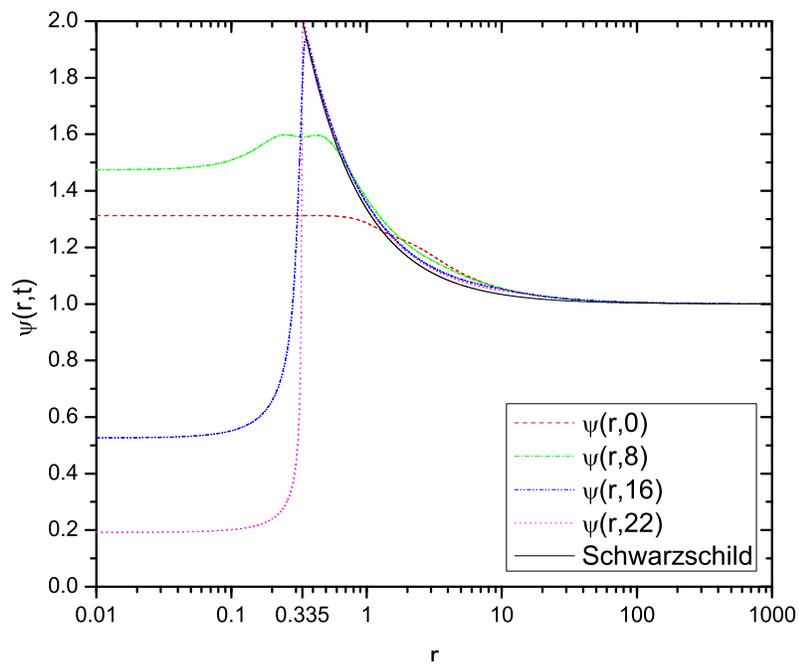}
		\end{center}
		\caption{ The conformal factor $\psi$  as a function of $r$ at different times. The solid line represents the exterior Schwarzschild analytical solution.}
		\label{psi}
\end{figure}

The metric for the static Schwarzschild vacuum exterior in isotropic coordinates is known analytically and is given by \reff{isotropic}. The numerical simulation should reproduce this metric in the exterior region and this provides a test on the accuracy of our code. In particular, the conformal factor in the exterior should have a form close to $\psi=1+M/(2r)$ and its peak value should be close to $2$. The theoretical and numerical curves for $\psi$ are shown in figure~\ref{psi}. Note how the numerical curves approach the analytical curve with time. At late times, the percentage difference between the two curves at any point is always under $1.31\%$, being significantly under this value most of the time. The peak value of $\psi$ analytically is equal to the integer $2$. For the run mentioned above we obtain a peak numerical value of $1.974$, when the ADM mass deviated by $1\%$ (at $t=18.28$) and a peak value of $\psi$ of $1.992$ when the ADM mass deviates by $5\%$ (at a later time $t= 20.26$). This corresponds to a tiny difference of $1.3\%$ and $0.4\%$ respectively with the expected analytical value. Note that the code is still evolving well (still approaching the analytical curve) from $t=18.28$ to $t=20.26$ even though the ADM mass has started to deviate from its original value. We ran the code for the following grid sizes: $\Delta x= \{10^{-3}, 5\times 10^{-4}, 2\times 10^{-4}, 1\times 10^{-4}, 5 \times 10^{-5}\}$. Let $\psi_1$ and $\psi_2$ be the peak values of $\psi$ at ADM mass deviation of $1\%$ and $5\%$ respectively. Their values for the corresponding grid sizes are $\psi_1=\{1.783, 1.859, 1.923, 1.953, 1.974\}$ and $\psi_2=\{1.897, 1.934, 1.966, 1.982, 1.992\}$. In both cases the values are converging and approaching closer to the expected analytical value of $2$.    

We also ran the code with different initial states: different values of $\lambda$ for the initial state of the form \reff{4DScalarIEChi} and a different initial state of ``non-Gaussian" form. The results are not very different and the features of all graphs remain the same. At highest resolution, we obtained $\psi_1=1.970$ and $\psi_2=1.989$ for $\lambda=1.55$ and $\psi_1=1.966$ and $\psi_2=1.986$ for $\lambda=1.60$. For a ``non-Gaussian" initial state of the form $\chi=\frac{\lambda_1}{1+\lambda_2 r^4}$ with $\lambda_1=0.3$ amd $\lambda_2=0.05$ we obtained $\psi_1=1.972$ and $\psi_2=1.978$, again at highest resolution. 

\subsection{Thin shell structure}

\begin{figure}[tbp]
		\begin{center}
			\includegraphics[scale=.50, draft=false, trim=2cm 1.5cm 2.5cm 2cm, clip=true]{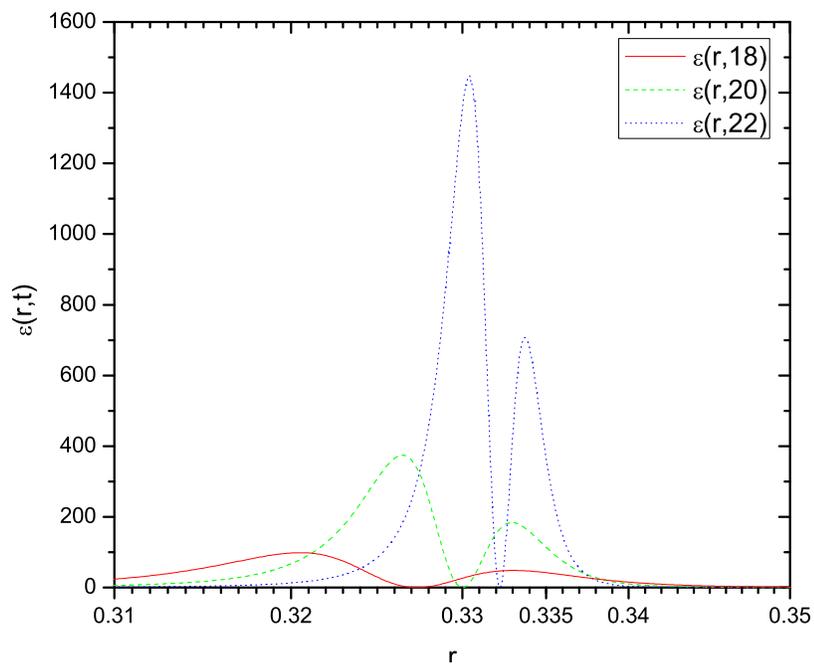}
		\end{center}
		\caption{The energy density $\mathcal{E}$ as a function of $r$ at different times.}
		\label{epsilon}
\end{figure}

\begin{figure}[tbp]
		\begin{center}
			\includegraphics[scale=.50, draft=false, trim=2cm 1.5cm 2.5cm 2cm, clip=true]{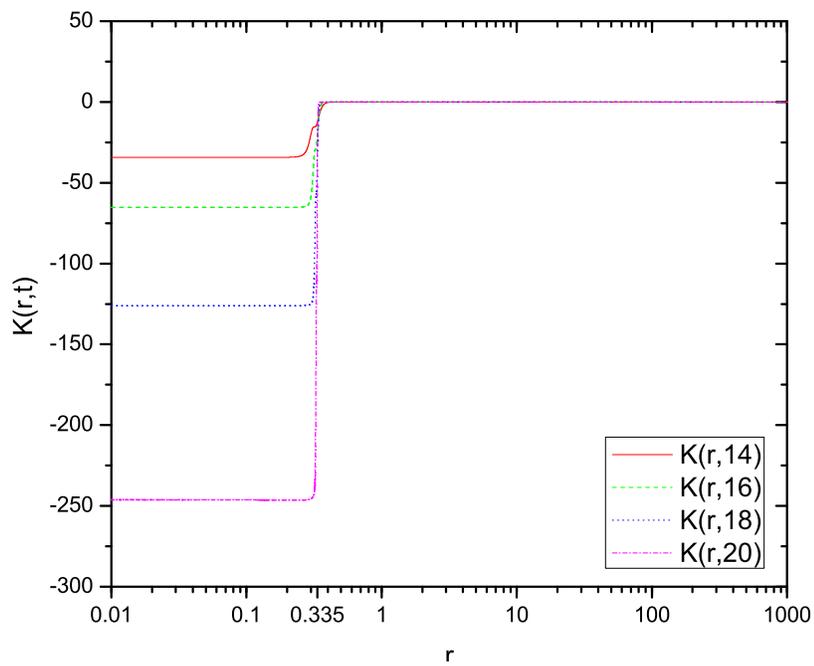}
		\end{center}
		\caption{The extrinsic curvature $K$ as a function of $r$ at different times. The jump discontinuity is due to the presence of a thin shell of matter near $r_h=0.335$.}
		\label{K}
\end{figure}

\begin{figure}[tbp]
		\begin{center}
			\includegraphics[scale=.50, draft=false, trim=2cm 1.5cm 2.5cm 2cm, clip=true]{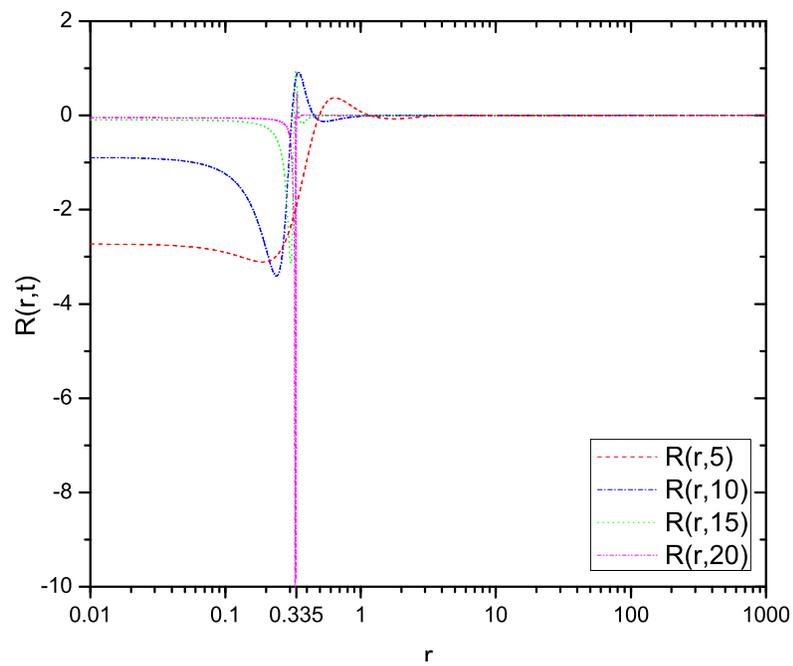}
		\end{center}
		\caption{The Ricci scalar $R$ as a function of $r$ at different times. At late times $R$ is basically zero everywhere except for a spike near the apparent horizon.}
		\label{R}
\end{figure}

\begin{figure}[tbp]
		\begin{center}
			\includegraphics[scale=.50, draft=false, trim=2cm 1.5cm 2.5cm 2cm, clip=true]{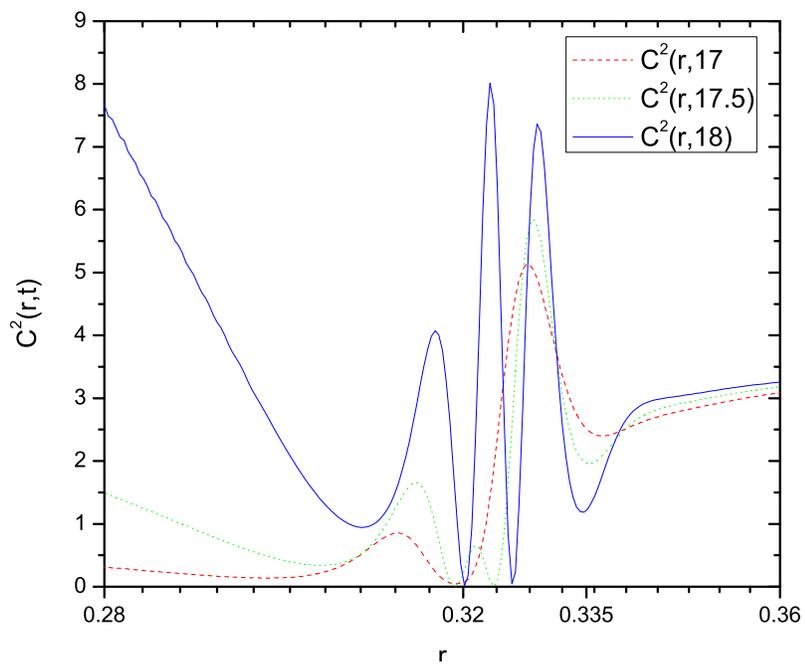}
		\end{center}
		\caption{ The Weyl tensor squared $C^2$ as a function of $r$ at different times with a focus on the thin shell region near $r_h$.}
		\label{Weyl2}
\end{figure}

At late times, the energy density is concentrated near two peaks\footnote{Recall that the initial scalar field configuration \reff{4DScalarIEChi} has an initial energy density profile with two peaks. The two peaks remain throughout the evolution.} just inside the apparent horizon, the $N\eq 0$ spacelike two-surface at $r_h=0.335$ (see fig.~\ref{epsilon}). During the collapse, the two peaks approach each other and the energy density profile becomes thinner with time so that the two effectively merge into one thin shell located just inside the horizon. The trace of the extrinsic curvature, $K$, takes the form of a Heaviside function as shown in fig.~\ref{K}, being very close to zero outside the horizon and being negative and spatially constant inside. The jump discontinuity in the extrinsic curvature is precisely what is expected from the presence of a thin shell of matter at the location where the jump occurs \cite{Poisson}. An ideal thin shell is expected to have a delta-function singularity in the stress-energy tensor $T^{\mu\nu}$ and the Riemann tensor $R^{\mu}_{\nu\sigma\tau}$ \cite{Poisson}. In our numerical simulation, these singularities are represented by various quantities increasing with time and reaching peak values or spikes at late times. We have already looked at the energy density (fig.~\ref{epsilon}), which is basically zero everywhere except for spikes located just inside the horizon and whose peak value increases with time. The Ricci scalar behaves in roughly the same fashion: it is zero everywhere except for spikes which increase in magnitude with time in the thin shell region (see fig.~\ref{R}). The square of the Weyl tensor ($C^2=C^{\mu\nu\sigma\tau}C_{\mu\nu\sigma\tau}$) has a few peaks just inside the horizon which can be seen to increase significantly with time (see fig.~\ref{Weyl2}). The spacetime can therefore be said to be evolving towards a curvature singularity at the thin shell (note that this occurs as the shell ``falls" towards the origin. See paragraph below). There is both a Ricci singularity associated with a spike in the energy density (and the Ricci scalar) and a Weyl singularity associated with a spike in $C^2$ representing strong tidal forces. Note that the standard Schwarzschild spacetime has a Weyl singularity but no Ricci singularity, since it is a purely vacuum solution. In the prototypical Oppenheimer-Snyder collapse \cite{Oppenheimer}, modeled after a contracting FLRW universe, the curvature singularity is due to a high energy density; there is a Ricci singularity but no Weyl singularity since the Weyl tensor is identically zero for an FLRW universe. Our collapse scenario leads to a more physically realistic situation where there is both a Ricci and Weyl curvature singularity. 

We will see shortly that the interior is a collapsing universe where $\psi$ is almost homogeneous and approaches zero. The thin shell is therefore ``falling" towards or approaching $r\eq 0$ while remaining at a radial coordinate near $r_h$. The proper radial distance between $r=0$ and $r=r_h$ is given by $\int_0^{r_h}\psi^2(r,t)\,dr$ which is approximately equal to $\psi^2(t) \,r_h$ (since $\psi$ is approximately homogeneous in the interior). This tends towards zero as $\psi$ approaches $0$ in the interior. The curvature scalars evolve towards the singularity as the shell ``falls" towards the origin.  

\subsection{Interior spacetime: a collapsing isotropic vacuum universe} 

\begin{figure}[tbp]
		\begin{center}
			\includegraphics[scale=.50, draft=false, trim=2cm 1.5cm 2.5cm 2cm, clip=true]{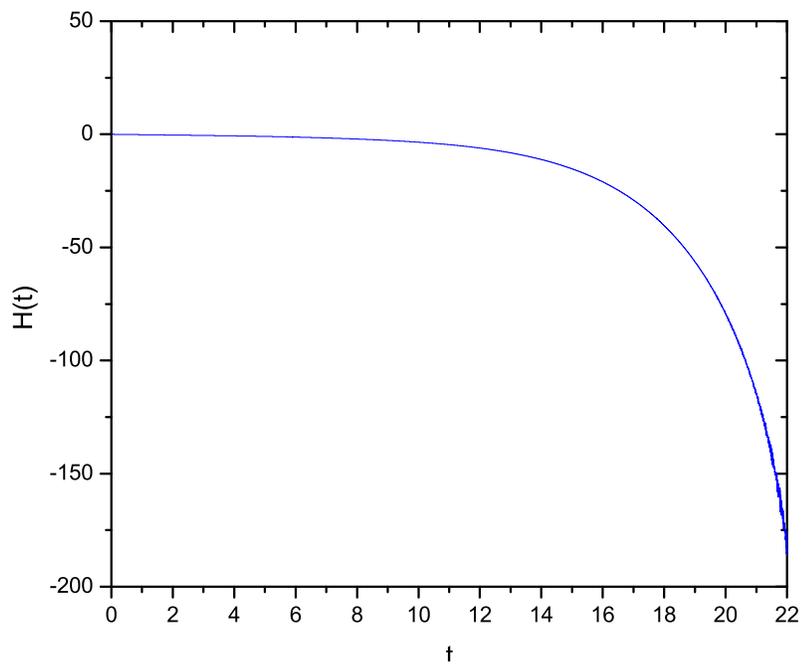}
		\end{center}
		\caption{The Hubble parameter as a function of time. The data is taken at $r=0$ but $H$ has almost no dependence on $r$.}
		\label{Hubble}
\end{figure}

\begin{figure}[tbp]
		\begin{center}
			\includegraphics[scale=.50, draft=false, trim=2cm 1.5cm 2.5cm 2cm, clip=true]{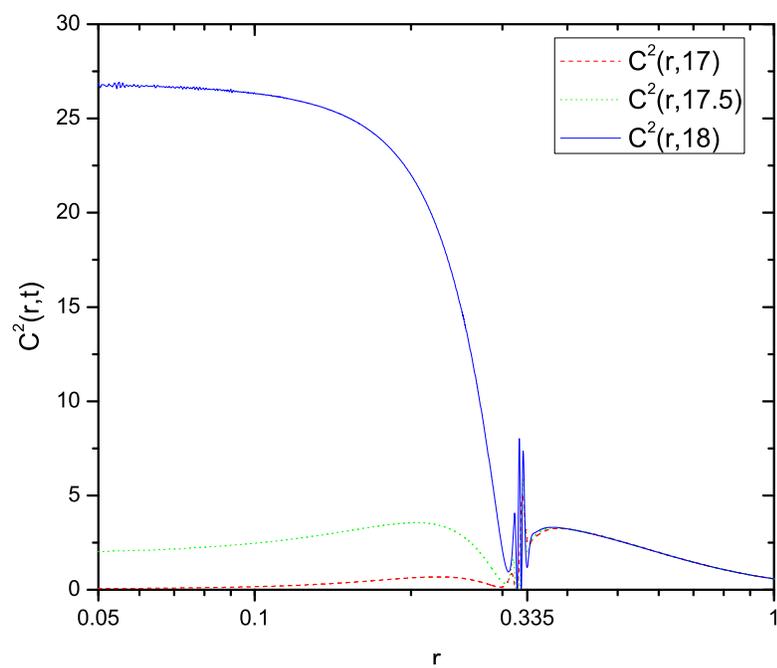}
		\end{center}
		\caption{The Weyl tensor squared $C^2$ as a function of $r$ at different times with a focus on the interior region (smaller $r$ values). Its value increases significantly in a very short time interval.}
		\label{Weyl1}
\end{figure}

Besides the emergence of a thin shell structure, the other salient feature of the spacetime is the interior. The interior is defined here as the nearly zero energy density (almost vacuum) region inside the apparent horizon where the thin shell of matter is absent. In the interior $\psi$ and $N$ are close to being homogeneous (see figures \ref{N} and \ref{psi}) so that $\psi(r,t)\approx \psi(t)$ and $N(r,t)\approx N(t)$. The magnitude of $\psi$ and $N$ are also decreasing with time towards zero: $N$ is approaching zero from below and $\psi$ from above\footnote{In the interior, $N$ and $\psi$ evolve towards zero but are never identically equal to zero.}.  The interior can be viewed {\it approximately} as a collapsing FLRW universe (approximately because the interior is not perfectly homogeneous but has a slight inhomogeneity). Defining $\tau=-\int N(t) dt$ the metric can be cast approximately in the FLRW form with flat spatial sections: $ds^2 \approx -d\tau^2 +\psi^4(\tau)(dr^2 + r^2 \,d\Omega^2)$\footnote{The collapse takes an infinite time $t$ but only a finite proper time $\tau$.}. The point of casting the metric in this approximate form is to extract the Hubble parameter as a measure of the rate of collapse. The scale factor is $a(\tau)=\psi^2(\tau)$ and the Hubble parameter is $H=\tfrac{1}{a}\tfrac{da}{d\tau}= \tfrac{2}{\psi}\tfrac{d\psi}{d\tau} = -\tfrac{2}{\psi}\tfrac{\dot{\psi}}{N}$ where a dot represents derivative with respect to the original $t$ coordinate. The Hubble parameter is plotted as a function of time in figure~\ref{Hubble}. It increases rapidly in magnitude with time and is negative since the interior is collapsing or contracting. The interior is clearly heading towards a curvature singularity as $\psi \to 0$. The areal radius $\psi^2\,r$ approaches zero inside and the Weyl tensor squared ($C^2$) in the interior region increases rapidly with time (see fig. ~\ref{Weyl1}). 

The trace of the extrinsic curvature $K=-6\dot{\psi}/(N\,\psi)$ can be expressed in terms of the above Hubble parameter as $K=3\,H$. The energy constraint equation \reff{E_Constraint} then takes the form
\beq
H^2=\dfrac{8\pi\,G}{3}\mathcal{E} +\frac{4}{3\,\psi^5}\,\nabla^2\psi \,.
\eeq{Hubble2}
In a perfectly homogeneous spacetime as in FLRW, $\nabla^2\psi$ would be zero and the above would reduce to the Friedmann equation for flat spatial sections and zero cosmological constant: $H^2=\tfrac{8\pi\,G}{3}\mathcal{E}$ where $\mathcal{E}$ is the energy density of the matter. Our interior is not exactly FLRW: it has a slight inhomogeneity which implies that $\nabla^2\psi\ne 0$. Also, the interior tends towards a vacuum: $\mathcal{E}\to 0$. For the interior, the governing equation is therefore $H^2=\tfrac{4}{3\,\psi^5}\,\nabla^2\psi$. We have seen that the magnitude of the Hubble parameter increases rapidly with time. Though there is only a slight inhomogeneity in $\psi$ so that $\nabla^2\psi$ is small, $\psi$ tends to zero in the interior so that the right hand side (RHS) term, which contains $\psi^5$ in the denominator, increases rapidly with time also. The RHS term also has a negligible spatial dependence like $K$ and $H$.  

In the exact FLRW case governed by the above Friedmann equation, the energy density eventually diverges as in Oppenheimer-Snyder collapse \cite{Oppenheimer}. In our slightly inhomogeneous contracting interior vacuum spacetime, it is the term $\frac{4}{3\,\psi^5}\,\nabla^2\psi$ that eventually diverges as $\psi \to 0$. The slight inhomogeneity plays a crucial role in the collapse process. The curvature singularity in the exact FLRW case is a Ricci singularity (due to high energy density). In contrast, our collapsing interior vacuum spacetime can be said to be evolving towards a Weyl curvature singularity, the same type of singularity encountered in the Schwarzschild vacuum interior. 

\section{Conclusion}
 
We studied numerically the spacetime which emerges after the collapse of a massless scalar field in isotropic coordinates. A salient feature of the spacetime is the emergence of a thin shell of matter just inside the horizon. The spacetime consists of two vacuum regions separated by the thin shell with the Ricci scalar being zero everywhere except for a spike at the shell. There is a jump discontinuity in the extrinsic curvature across the horizon, in accord with the presence of a thin shell near the horizon. The expected delta function singularities at an ideal thin shell \cite{Poisson} are represented numerically as spikes in the Ricci scalar, the energy density and the Weyl tensor squared which all increase significantly with time in the thin shell region. The exterior region settles into the static Schwarzschild vacuum solution in isotropic coordinates. The interior is an isotropic universe that collapses basically in vacuo. The magnitude of the ``Hubble parameter" and the trace of the extrinsic curvature increase with time not because the energy density increases, but because a term proportional to $\nabla^2\psi/\psi^5$, which has a non-zero value due to a slight inhomogeneity, grows as $\psi$ tends to zero. The Weyl tensor squared rises rapidly with time in the interior as the areal radius shrinks. 

The most natural case to study next would be charged collapse in isotropic coordinates. There are many interesting questions to explore. Would the charge density be concentrated in a thin shell and have a similar profile as the energy density? Would there be a timelike singularity and inner horizon as in the Reissner-Nordstr\"{o}m (RN) spacetime? This question is important because the inner horizon in RN is thought to be unstable \cite{Poisson} and charged collapse studies in other coordinates have not revealed a timelike singularity \cite{Brady,Piran}. But more importantly, isotropic coordinates would allow one to study the thermodynamics of the charged black hole at late stages of the collapse. In particular, one could track the gravitational Lagrangian, a function which has been shown to be connected to thermodynamic properties of a black hole \cite{Khlebnikov,Benjamin-Edery2,Edery_Constantineau}.

\section*{Acknowledgments}
A.E. acknowledges support from a discovery grant of the National
Science and Engineering Research Council of Canada (NSERC). H.B. 
acknowledges financial support from a Bishop's Senate Research Grant.

\end{document}